\documentstyle[tighten,preprint,amsfonts,eqsecnum,aps,prd,bezier]{revtex}
\begin{document}
\draft

\hyphenation{
mani-fold
mani-folds
}



\def\BbbR{{\Bbb R}}
\def\BbbZ{{\Bbb Z}}

\def\etwo{{E^2}}
\def\othree{{O(3)}}
\def\sothree{{SO(3)}}
\def\sotwoone{{SO(2,1)}}
\def\soctwoone{{SO_c(2,1)}}

\def\RPtwo{{\BbbR P^2}}
\def\RPthree{{\BbbR P^3}}

\def\ads{{anti-de~Sitter}}
\def\rnads{{Reissner-Nordstr\"om-anti-de~Sitter}}

\def\casehalf{{\case{1}{2}}}

\def\binfty{\beta_\infty}

\def\bm{{\bf m}}
\def\bq{{\bf q}}

\def\Rh{R_{\rm h}}
\def\Rnought{{\bf R}_{\rm h}}
\def\Rhor{R_{\rm hor}}
\def\Rcrit{R_{\rm crit}}
\def\Mcrit{M_{\rm crit}}


\preprint{\vbox{\baselineskip=12pt
\rightline{PP97--111}
\rightline{USITP 97--6}
\rightline{gr-qc/9705012}}}
\title{Thermodynamics of ($3+1$)-dimensional black holes with
\\
toroidal or higher genus horizons}
\author{Dieter R. Brill\footnote{Electronic address:
brill@umdhep.umd.edu}
and
Jorma Louko\footnote{On leave of absence from 
Department of Physics, University of Helsinki.
Electronic address:
louko@wam.umd.edu~.
Address after September~1, 1997:
Max-Planck-Institut f\"ur Gravitations\-physik,
Schlaatzweg~1,
D--14473 Potsdam,
Germany.
}}
\address{
Department of Physics,
University of Maryland,
College Park,
Maryland 20742--4111,
USA}
\author{Peter Peld\'an\footnote{Electronic address:
peldan@vanosf.physto.se}}
\address{
Fysikum,
Stockholm University,
Box 6730,
S-113~85 Stockholm,
Sweden}
\bigskip
\date{Revised version, July 1997. To be published in Phys.\ Rev. D.}
\maketitle
\begin{abstract}%
We examine counterparts of the \rnads\ black hole spacetimes in which
the two-sphere has been replaced by a surface $\Sigma$ of constant
negative or zero curvature. When horizons exist, the spacetimes are black
holes with an asymptotically locally \ads\ infinity, but the infinity
topology differs from that in the asymptotically Minkowski case, and the
horizon topology is not~$S^2$. Maximal analytic extensions of the solutions
are given. The local Hawking temperature is found. When $\Sigma$ is
closed, we derive the first law of thermodynamics using a Brown-York type
quasilocal energy at a finite boundary, and we identify the entropy as
one quarter of the horizon area, independent of the horizon topology. The
heat capacities with constant charge and constant electrostatic potential
are shown to be positive definite. With the boundary pushed to infinity, we
consider thermodynamical ensembles that fix the renormalized temperature
and either the charge or the electrostatic potential at infinity. Both
ensembles turn out to be thermodynamically stable, and dominated by a
unique classical solution.
\end{abstract}

\narrowtext

\section{Introduction}
\label{sec:intro}

Isolated black holes created in astrophysical processes are expected to be
well described by Einstein spacetimes that are asymptotic to Minkowski space
near a spacelike or null infinity. A familiar example is the Kerr-Newman
family of Einstein-Maxwell black holes \cite{MTW}. However, there is
mathematical interest in black holes with other kinds of asymptotic
infinities. One alternative is to consider black holes that are
asymptotically \ads\ in the sense of Refs.\
\cite{abb-deser,Ash-ads,HennTeit-ads},
so that the topology at infinity is
the same as that in asymptotically flat spacetimes. An example of this in
four spacetime dimensions is the Kerr-Newman-\ads\ black hole family
\cite{carter-cmp,carter-les,lake1}, which generalizes the Kerr-Newman
family to accommodate a negative cosmological constant. Examples in other
dimensions include the Ba\~{n}ados-Teitelboim-Zanelli (BTZ) black hole
\cite{btz-hole,carlip-rev} and its dimensionally continued relatives
\cite{btz-cont}.

In this paper we examine a class of four-dimensional black holes
that are asymptotically \ads, but whose topology near infinity differs
from that in the asymptotically Minkowski case. These spacetimes solve the
Einstein-Maxwell equations with a negative cosmological constant: they
generalize the \rnads\ solutions, replacing the round two-sphere by a
two-dimensional space $\Sigma$ of constant negative or vanishing curvature.
These spacetimes emerge as the generic solution family from a sufficiently
general form of Birkhoff's theorem, and their local geometry is well
understood \cite{exact-book,bronnikov}.
The purpose of the present paper is
to examine the global structure of these spacetimes appropriate for a black
hole interpretation, and the thermodynamics of the black hole spacetimes.
In particular, we shall address the thermodynamical stability of these
black holes under suitable boundary conditions, both with a finite
boundary and with an asymptotic infinity. Our results generalize those
obtained previously in Refs.\
\cite{lemos-plan,lemos-cyl,huang,lemos-zan,cai,%
ABHP,mann-instanton,mann-smith,banados4,mann-cloud}.
Preliminary results were briefly mentioned in Refs.\
\cite{ABHP,brill-multi2,brill-multi3}.

We begin, in section~\ref{sec:spacetimes}, by describing the local and
global structure of the spacetimes. All the spacetimes have
one or more asymptotically \ads\ infinities, and we can use Killing time
translations at infinity to define Arnowitt-Deser-Misner (ADM) mass and
charge. These quantities turn out to be finite if $\Sigma$ is closed. The
number and character of the Killing horizons depends on the 
parameters in the metric. 
Whenever a nondegenerate Killing horizon exists, the spacetime has an
interpretation as a black hole, and the (outer) Killing horizon has an
interpretation as a black hole horizon. The (outer) Killing
horizon bifurcation two-space has the topology of~$\Sigma$.
If the additive constant in the ADM mass is chosen so that this mass
vanishes for the solutions that are locally \ads, we find that black holes
with flat $\Sigma$ necessarily have positive ADM mass, but when $\Sigma$ has
negative curvature, there are black hole solutions with either sign of the
ADM mass. The spacetimes with a degenerate Killing horizon are not black
holes, in contrast to (say) the extreme Reissner-Nordstr\"om black hole
\cite{haw-ell}; the reason for this difference is that the negative
cosmological constant makes the future null infinity in our spacetimes
connected.

Section \ref{sec:finite-thermo} addresses the thermodynamics of the
black hole spacetimes. The local Hawking temperature is found from the Unruh
effect, or from the periodicity of Euclidean time, in terms of the
surface gravity at the horizon. Taking $\Sigma$ closed, we introduce a
boundary with the topology of $\Sigma$ and fixed size, and we find the
Brown-York type quasilocal energy at this boundary. Interpreting this
quasilocal energy as the internal thermodynamical energy, and using the
local Hawking temperature, we write the first law of black hole
thermodynamics. We find that for all the horizon topologies, the entropy is
one quarter of the horizon area. This result extends the
Bekenstein-Hawking area law to toroidal and higher genus horizons. In the
limit of a large box, we show that the heat capacities with fixed ADM
charge and fixed electrostatic potential are always positive.

In section \ref{sec:infinite-thermo} we consider the thermodynamics in the
limit where the boundary is pushed strictly to infinity. As the local
Hawking temperature vanishes at infinity, we focus on the
renormalized temperature that is obtained by multiplying the local
temperature by the redshift factor. As with the conventional \rnads\ black
holes \cite{HPads,PagePhill,pagerev,BrownCreMann,lou-win}, this turns out
to yield a first law from which the entropy emerges as one quarter of the
horizon area. We consider the canonical ensemble, in which one fixes the ADM
charge, and the grand canonical ensemble, in which one fixes the
electrostatic potential difference between the horizon and the infinity
with respect to the Killing time. The (path) integral expression for the
(grand) partition function is obtained by adapting to our symmetries the
Hamiltonian reduction techniques of Refs.\
\cite{lou-win,WYprl,whitingCQG,BBWY,LW2,BLPP,LouSiWi}. Both ensembles turn
out to be thermodynamically stable, and always dominated by a unique
classical black hole solution.

Section \ref{sec:discussion} contains a brief summary and discussion.
Some of the technical detail on the heat capacities is collected in the
appendix.

We work throughout in Planck units, $\hbar = c = G = 1$.

\section{Black hole spacetimes}
\label{sec:spacetimes}

\subsection{Local curvature properties}
\label{subsec:local-curvature}

We consider spacetimes whose metric can be written locally in the form
\begin{mathletters}
\label{gen-metric}
\begin{equation}
ds^2 = - F dT^2 + F^{-1} dR^2 + R^2 d\Omega_k^2
\ \ ,
\end{equation}
where
\begin{equation}
F := k - {2M \over R} + {Q^2 \over R^2} - {\Lambda R^2 \over 3}
\ \ .
\label{F}
\end{equation}
\end{mathletters}%
The parameters $M$, $Q$, and $\Lambda$ are real and continuous. The
discrete parameter $k$ takes the values~$1$, $0$, and~$-1$, and
$d\Omega_k^2$ is the metric on a two-dimensional surface $\Sigma_k$ of
constant Gaussian curvature~$k$.
In local coordinates $(\theta,\varphi)$
on~$\Sigma_k$, we can write
\begin{equation}
d\Omega_k^2 = \cases{%
d\theta^2 + \sin^2(\theta) \, d\varphi^2\ ,&$k=1$;\cr
d\theta^2 + \theta^2 d\varphi^2\ ,&$k=0$;\cr
d\theta^2 + \sinh^2 (\theta) \, d\varphi^2\ ,&$k=-1$. \cr}
\end{equation}
$\Sigma_k$ is locally homogeneous \cite{wolf,ash-sam},
with the local isometry group $\sothree$ for $k=1$, $\etwo$ for $k=0$, and
$\soctwoone$ (the connected component of $\sotwoone$) for $k=-1$.
The local isometries of $\Sigma_k$
are clearly inherited by the four-dimensional metric~(\ref{gen-metric}). The
vector $\partial/\partial{T}$ is a Killing vector, timelike for $F>0$ and
spacelike for $F<0$. We refer to $T$ as the Killing time, and to the
coordinates $(T,R)$ as the curvature coordinates. Without loss of
generality, we can assume $R>0$.

The metric (\ref{gen-metric}) solves the Einstein-Maxwell equations with
the cosmological constant $\Lambda$ and the electromagnetic potential
one-form
\begin{equation}
\bbox{A} = {Q \over R} \bbox{d}T
\ \ .
\label{A}
\end{equation}
Indeed, the metric (\ref{gen-metric}) with the electromagnetic
potential (\ref{A}) emerges from a sufficiently general form of
Birkhoff's theorem as the generic family of Einstein-Maxwell spacetimes
admitting the local isometry group $\sothree$, $\etwo$, or $\soctwoone$
with two-dimensional spacelike orbits \cite{exact-book}. Our electromagnetic
potential (\ref{A}) yields a vanishing magnetic field, but the spacetimes
with a nonvanishing magnetic field can be obtained from (\ref{A}) by the
electromagnetic duality rotation.

\subsection{Global properties}
\label{subsec:global-properties}

We now examine the global properties of the spacetimes
(\ref{gen-metric}) with $\Lambda<0$. We write $\Lambda = -3
\ell^{-2}$ with $\ell>0$.

The first issue is in the global geometry of~$\Sigma_k$. To exclude
spacetime singularities that would result solely from singularities in the
two-dimensional geometry of~$\Sigma_k$, we take $\Sigma_k$ to be complete.
We can then write $\Sigma_k={\tilde\Sigma}_k/\Gamma$, where
${\tilde\Sigma}_k$ is the universal covering space of~$\Sigma_k$,
and $\Gamma$ is a freely and properly discontinuously
acting subgroup of the full
isometry group of~${\tilde\Sigma}_k$.
If the action of $\Gamma$ on ${\tilde\Sigma}_k$ is
nontrivial, $\Sigma_k$ is multiply connected.

For $k=1$, ${\tilde\Sigma}_1$ is $S^2$ with the round metric. The
isometry group is~$\othree$. The only multiply connected choice for
$\Sigma_1$ is $\RPtwo=S^2/\BbbZ_2$, where the nontrivial element of
$\BbbZ_2$ is the antipodal map \cite{wolf}.

For $k=0$, ${\tilde\Sigma}_0$ is $\BbbR^2$ with the flat metric.
The isometry group is $\etwo \times_s \BbbZ_2$, where the nontrivial element
of $\BbbZ_2$ is the reflection about a prescribed geodesic, and $\times_s$
stands for the semidirect product. The multiply connected choices for
$\Sigma_0$ are the cylinder, the M\"obius band, the torus, and the Klein
bottle \cite{wolf}.

For $k=-1$, ${\tilde\Sigma}_{-1}$ is $\BbbR^2$ with the hyperbolic
metric. The isometry group is $\soctwoone \times_s \BbbZ_2$, where the
nontrivial element of $\BbbZ_2$ is the reflection about a prescribed
geodesic.  The closed and orientable choices for $\Sigma_{-1}$ are
the closed Riemann surfaces of genus $g>1$ (see for example Ref.\
\cite{nag}). The multiply connected but not closed choices for
$\Sigma_{-1}$ include the cylinder \cite{ABHP,banados4} and the M\"obius
band, as well as surfaces with an arbitrary finite number of infinities
\cite{brill-multi2,brill-multi3,brill-multi1}.

When $\Sigma_{k}$ is closed, we denote its area by~$V$. For
$k=1$, both $S^2$ and $\RPtwo$ are closed, and we have respectively
$V=4\pi$ and $V=2\pi$. For $k=0$, the closed choices are the torus and the
Klein bottle, and $V$ can in either case take arbitrary positive values.
For $k=-1$, with $\Sigma_{-1}$ closed, the Gauss-Bonnet theorem (see for
example Refs.\
\cite{spivak,allen-weil}) implies $V= -2\pi\chi$, where $\chi$ is the
Euler number of $\Sigma_{-1}$, and $V$ is therefore completely determined
by the topology. In the orientable case, we have $\chi = 2(1 - g)$ and $V
= 4\pi(g-1)$.

We next turn to the infinity structure of the metric~(\ref{gen-metric}).
At $R\to\infty$, the dominant behavior of the metric is determined by
the cosmological constant for any values of $M$ and~$Q$. In the special
case $M=0=Q$, the spacetime is locally isometric to \ads\
space \cite{btz-cont,ABHP,banados4}.
We can therefore regard the infinity at $R\to\infty$ as an
asymptotically locally \ads\ infinity for any values of $M$ and~$Q$. The
precise sense of this asymptotic structure has been examined in Refs.\
\cite{Ash-ads,HennTeit-ads,btz-cont,lou-win} for $k=1$, and the
Hamiltonian falloff analyses of Refs.\ \cite{btz-cont,lou-win} can be
readily adapted to cover also the cases $k=0$ and $k=-1$. The infinity
is both a spacelike and a null infinity. In a Penrose diagram that
suppresses~$\Sigma_k$, the infinity can be represented by a vertical line.

For $k=1$ and $\Sigma_{1}=S^2$, the asymptotic \ads\ symmetry at
$R\to\infty$ allows one to introduce a Hamiltonian formulation with a
well-defined Arnowitt-Deser-Misner (ADM) Hamiltonian
\cite{Ash-ads,HennTeit-ads,btz-cont,lou-win}. This Hamiltonian generates
translations of the spacelike hypersurfaces at infinity with respect to
the asymptotic Killing time, normalized as the coordinate
$T$ in~(\ref{gen-metric}). It is straightforward to adapt the techniques
of Refs.\ \cite{btz-cont,lou-win} to show that the same conclusion holds
for all of our metrics for which $\Sigma_k$ is closed. If one normalizes
the additive constant in the Hamiltonian so that the Hamiltonian vanishes
for $M=0=Q$, one finds that the contribution of an infinity to the ADM
Hamiltonian is~$(V/4\pi)M$, and the contribution to the analogously defined
ADM electric charge is~$(V/4\pi)Q$.
When $\Sigma_k$ is not closed, however,
the infinite area of $\Sigma_k$ implies infinite values for both the
Hamiltonian and the charge.

Consider next the singularity structure of the metric~(\ref{gen-metric}).
The metric has a curvature singularity at $R\to0$ except when
$M=0=Q$. When $M=0=Q$, the spacetime is locally \ads, and the behavior at
$R\to0$ depends on the topology of $\Sigma_k$. If $\Sigma_k$ is simply
connected, the spacetime (\ref{gen-metric}) with $R>0$ is isometric to a
certain region of \ads\ space \cite{btz-cont,ABHP,banados4}: $R\to0$ is
then a mere coordinate singularity, and the spacetime can be continued
past $R=0$ to all of \ads\ space.  If $\Sigma_k$ is not simply connected,
the spacetime (\ref{gen-metric}) with $R>0$ is isometric to a quotient
space of a certain region of \ads\ space with respect to
a discrete subgroup of the isometry group, and the possibilities of
continuing the spacetime past $R=0$ depend on how these discrete
isometries extend to the rest of \ads\ space. Typically, the extended
spacetime is singular in its topological structure \cite{ABHP,banados4}, in
analogy with Misner space \cite{haw-ell} or the BTZ black hole
\cite{btz-hole,carlip-rev}.  We shall
not attempt to classify these singularities here.

We can now turn to the horizon structure. As usual \cite{walker}, the
positive values of $R$ at which the function $F(R)$ [Eq.~(\ref{F})]
vanishes are coordinate singularities on null hypersurfaces. The vector
$\partial/\partial{T}$ is a globally-defined Killing vector, timelike in
the regions with $F>0$, spacelike in the regions with $F<0$, and null on
the hypersurfaces with $F=0$. The regions with $F>0$ are therefore static,
and the hypersurfaces with $F=0$ are Killing horizons.

For examining the (positive) zeroes of~$F(R)$, it is
useful to define the quantity
\begin{equation}
\Mcrit(Q) := {\ell \over 3 \sqrt{6} }
\left( \sqrt{k^2 + 12 {(Q/\ell)}^2} +2k \right)
{\left( \sqrt{k^2 + 12 {(Q/\ell)}^2} -k \right)}^{1/2}
\ \ .
\label{McritQ}
\end{equation}
For $k=1$, a complete analysis can be found in
Refs.\ \cite{lake1,btz-cont,lou-win}. We shall therefore from
now on only consider the cases $k=0$ and $k=-1$.

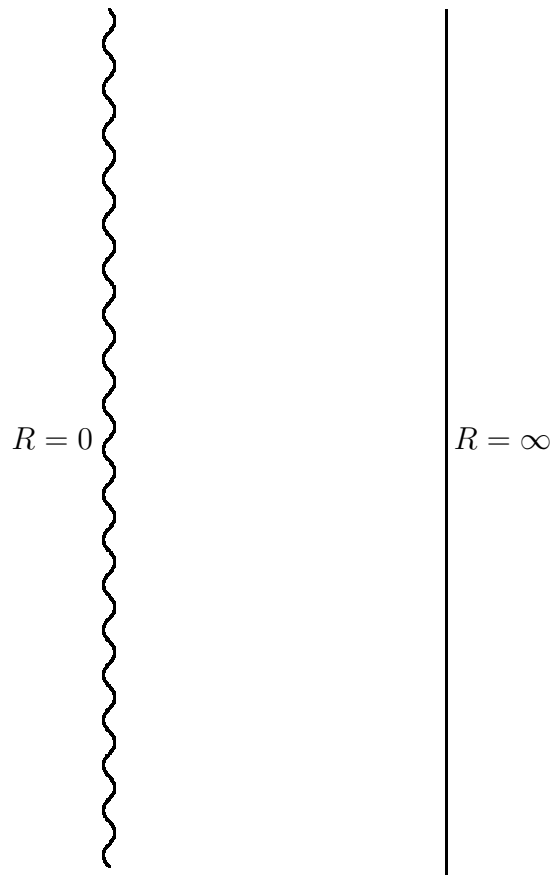
\begin{figure}
\vglue 1 cm

\unitlength 1.00mm
\linethickness{0.4pt}
\begin{picture}(106.00,122.00)
\thicklines
\put(105.00,7.00){\line(0,1){115.00}}
\bezier{20}(60.00,11.00)(61.50,12.50)(60.00,14.00)
\bezier{20}(60.00,17.00)(61.50,18.50)(60.00,20.00)
\bezier{20}(60.00,23.00)(61.50,24.50)(60.00,26.00)
\bezier{20}(60.00,29.00)(61.50,30.50)(60.00,32.00)
\bezier{20}(60.00,35.00)(61.50,36.50)(60.00,38.00)
\bezier{20}(60.00,41.00)(61.50,42.50)(60.00,44.00)
\bezier{20}(60.00,47.00)(61.50,48.50)(60.00,50.00)
\bezier{20}(60.00,53.00)(61.50,54.50)(60.00,56.00)
\bezier{20}(60.00,59.00)(61.50,60.50)(60.00,62.00)
\bezier{20}(60.00,8.00)(58.50,9.50)(60.00,11.00)
\bezier{20}(60.00,14.00)(58.50,15.50)(60.00,17.00)
\bezier{20}(60.00,20.00)(58.50,21.50)(60.00,23.00)
\bezier{20}(60.00,26.00)(58.50,27.50)(60.00,29.00)
\bezier{20}(60.00,32.00)(58.50,33.50)(60.00,35.00)
\bezier{20}(60.00,38.00)(58.50,39.50)(60.00,41.00)
\bezier{20}(60.00,44.00)(58.50,45.50)(60.00,47.00)
\bezier{20}(60.00,50.00)(58.50,51.50)(60.00,53.00)
\bezier{20}(60.00,56.00)(58.50,57.50)(60.00,59.00)
\bezier{20}(60.00,62.00)(58.50,63.50)(60.00,65.00)
\bezier{20}(60.00,65.00)(61.50,66.50)(60.00,68.00)
\bezier{20}(60.00,68.00)(58.50,69.50)(60.00,71.00)
\bezier{20}(60.00,71.00)(61.50,72.50)(60.00,74.00)
\bezier{20}(60.00,77.00)(61.50,78.50)(60.00,80.00)
\bezier{20}(60.00,83.00)(61.50,84.50)(60.00,86.00)
\bezier{20}(60.00,89.00)(61.50,90.50)(60.00,92.00)
\bezier{20}(60.00,95.00)(61.50,96.50)(60.00,98.00)
\bezier{20}(60.00,101.00)(61.50,102.50)(60.00,104.00)
\bezier{20}(60.00,107.00)(61.50,108.50)(60.00,110.00)
\bezier{20}(60.00,113.00)(61.50,114.50)(60.00,116.00)
\bezier{20}(60.00,119.00)(61.50,120.50)(60.00,122.00)
\bezier{20}(60.00,74.00)(58.50,75.50)(60.00,77.00)
\bezier{20}(60.00,80.00)(58.50,81.50)(60.00,83.00)
\bezier{20}(60.00,86.00)(58.50,87.50)(60.00,89.00)
\bezier{20}(60.00,92.00)(58.50,93.50)(60.00,95.00)
\bezier{20}(60.00,98.00)(58.50,99.50)(60.00,101.00)
\bezier{20}(60.00,104.00)(58.50,105.50)(60.00,107.00)
\bezier{20}(60.00,110.00)(58.50,111.50)(60.00,113.00)
\bezier{20}(60.00,116.00)(58.50,117.50)(60.00,119.00)
\thinlines
\put(106.00,65.00){\makebox(0,0)[lc]{$R=\infty$}}
\put(58.00,65.00){\makebox(0,0)[rc]{$R=0$}}
\end{picture}

\vglue 1 cm
\caption{The Penrose diagram for $M<\Mcrit$. The  straight line
indicates an infinity and the wavy line a singularity.}
\label{fig:q-penless}

\end{figure}

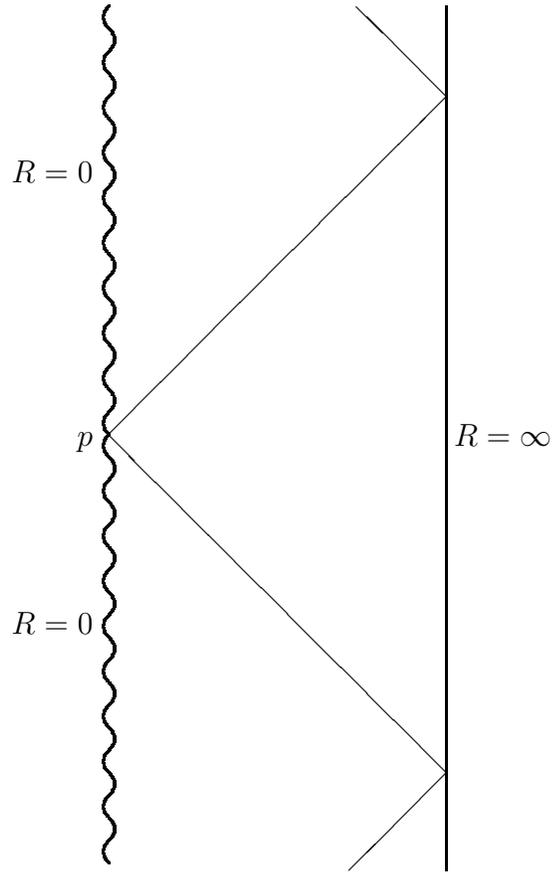
\begin{figure}
\vglue 1 cm

\unitlength 1.00mm
\linethickness{0.4pt}
\begin{picture}(106.00,122.00)
\put(105.00,20.00){\line(-1,1){45.00}}
\put(60.00,65.00){\line(1,1){45.00}}
\put(105.00,20.00){\line(-1,-1){13.00}}
\put(105.00,110.00){\line(-1,1){12.00}}
\thicklines
\put(105.00,7.00){\line(0,1){115.00}}
\bezier{20}(60.00,11.00)(61.50,12.50)(60.00,14.00)
\bezier{20}(60.00,17.00)(61.50,18.50)(60.00,20.00)
\bezier{20}(60.00,23.00)(61.50,24.50)(60.00,26.00)
\bezier{20}(60.00,29.00)(61.50,30.50)(60.00,32.00)
\bezier{20}(60.00,35.00)(61.50,36.50)(60.00,38.00)
\bezier{20}(60.00,41.00)(61.50,42.50)(60.00,44.00)
\bezier{20}(60.00,47.00)(61.50,48.50)(60.00,50.00)
\bezier{20}(60.00,53.00)(61.50,54.50)(60.00,56.00)
\bezier{20}(60.00,59.00)(61.50,60.50)(60.00,62.00)
\bezier{20}(60.00,8.00)(58.50,9.50)(60.00,11.00)
\bezier{20}(60.00,14.00)(58.50,15.50)(60.00,17.00)
\bezier{20}(60.00,20.00)(58.50,21.50)(60.00,23.00)
\bezier{20}(60.00,26.00)(58.50,27.50)(60.00,29.00)
\bezier{20}(60.00,32.00)(58.50,33.50)(60.00,35.00)
\bezier{20}(60.00,38.00)(58.50,39.50)(60.00,41.00)
\bezier{20}(60.00,44.00)(58.50,45.50)(60.00,47.00)
\bezier{20}(60.00,50.00)(58.50,51.50)(60.00,53.00)
\bezier{20}(60.00,56.00)(58.50,57.50)(60.00,59.00)
\bezier{20}(60.00,62.00)(58.50,63.50)(60.00,65.00)
\bezier{20}(60.00,119.00)(61.50,117.50)(60.00,116.00)
\bezier{20}(60.00,113.00)(61.50,111.50)(60.00,110.00)
\bezier{20}(60.00,107.00)(61.50,105.50)(60.00,104.00)
\bezier{20}(60.00,101.00)(61.50,99.50)(60.00,98.00)
\bezier{20}(60.00,95.00)(61.50,93.50)(60.00,92.00)
\bezier{20}(60.00,89.00)(61.50,87.50)(60.00,86.00)
\bezier{20}(60.00,83.00)(61.50,81.50)(60.00,80.00)
\bezier{20}(60.00,77.00)(61.50,75.50)(60.00,74.00)
\bezier{20}(60.00,71.00)(61.50,69.50)(60.00,68.00)
\bezier{20}(60.00,122.00)(58.50,120.50)(60.00,119.00)
\bezier{20}(60.00,116.00)(58.50,114.50)(60.00,113.00)
\bezier{20}(60.00,110.00)(58.50,108.50)(60.00,107.00)
\bezier{20}(60.00,104.00)(58.50,102.50)(60.00,101.00)
\bezier{20}(60.00,98.00)(58.50,96.50)(60.00,95.00)
\bezier{20}(60.00,92.00)(58.50,90.50)(60.00,89.00)
\bezier{20}(60.00,86.00)(58.50,84.50)(60.00,83.00)
\bezier{20}(60.00,80.00)(58.50,78.50)(60.00,77.00)
\bezier{20}(60.00,74.00)(58.50,72.50)(60.00,71.00)
\bezier{20}(60.00,68.00)(58.50,66.50)(60.00,65.00)
\thinlines
\put(58.00,64.00){\makebox(0,0)[rc]{$p$}}
\put(58.00,100.00){\makebox(0,0)[rc]{$R=0$}}
\put(58.00,40.00){\makebox(0,0)[rc]{$R=0$}}
\put(106.00,65.00){\makebox(0,0)[lc]{$R=\infty$}}
\end{picture}

\vglue 1 cm
\caption{The Penrose diagram for $M=\Mcrit$,
if $Q\ne0$ or $k=-1$ or both. The point $p$ is an internal
spacelike infinity, and the singularity consists of countably many
connected components. The infinity, which is both spacelike and (future)
null, consists of a single connected component. As the past of the
infinity consists of all of the spacetime, the spacetime does not have an
interpretation as a black hole.}
\label{fig:q-peneq}
\end{figure}

\begin{figure}
\vglue 1 cm

\unitlength 1.00mm 
\linethickness{0.4pt}
\begin{picture}(106.00,168.00)
\put(60.00,20.00){\line(1,1){45.00}}
\put(105.00,65.00){\line(-1,1){45.00}}
\put(60.00,110.00){\line(1,1){45.00}}
\put(105.00,155.00){\line(-1,1){13.00}}
\put(60.00,20.00){\line(1,-1){13.00}}
\put(105.00,20.00){\line(-1,1){45.00}}
\put(60.00,65.00){\line(1,1){45.00}}
\put(105.00,110.00){\line(-1,1){45.00}}
\put(60.00,155.00){\line(1,1){13.00}}
\put(105.00,20.00){\line(-1,-1){13.00}}
\thicklines
\bezier{20}(60.00,8.00)(61.50,9.50)(60.00,11.00)
\bezier{20}(60.00,11.00)(58.50,12.50)(60.00,14.00)
\bezier{20}(60.00,14.00)(61.50,15.50)(60.00,17.00)
\bezier{20}(60.00,17.00)(58.50,18.50)(60.00,20.00)
\put(60.00,20.00){\line(0,1){45.00}}
\bezier{20}(60.00,68.00)(61.50,69.50)(60.00,71.00)
\bezier{20}(60.00,74.00)(61.50,75.50)(60.00,77.00)
\bezier{20}(60.00,80.00)(61.50,81.50)(60.00,83.00)
\bezier{20}(60.00,86.00)(61.50,87.50)(60.00,89.00)
\bezier{20}(60.00,92.00)(61.50,93.50)(60.00,95.00)
\bezier{20}(60.00,98.00)(61.50,99.50)(60.00,101.00)
\bezier{20}(60.00,104.00)(61.50,105.50)(60.00,107.00)
\bezier{20}(60.00,65.00)(58.50,66.50)(60.00,68.00)
\bezier{20}(60.00,71.00)(58.50,72.50)(60.00,74.00)
\bezier{20}(60.00,77.00)(58.50,78.50)(60.00,80.00)
\bezier{20}(60.00,83.00)(58.50,84.50)(60.00,86.00)
\bezier{20}(60.00,89.00)(58.50,90.50)(60.00,92.00)
\bezier{20}(60.00,95.00)(58.50,96.50)(60.00,98.00)
\bezier{20}(60.00,101.00)(58.50,102.50)(60.00,104.00)
\bezier{20}(60.00,107.00)(58.50,108.50)(60.00,110.00)
\put(60.00,110.00){\line(0,1){45.00}}
\bezier{20}(60.00,158.00)(61.50,159.50)(60.00,161.00)
\bezier{20}(60.00,164.00)(61.50,165.50)(60.00,167.00)
\bezier{20}(60.00,155.00)(58.50,156.50)(60.00,158.00)
\bezier{20}(60.00,161.00)(58.50,162.50)(60.00,164.00)
\bezier{20}(105.00,8.00)(102.94,9.50)(105.00,11.00)
\bezier{20}(105.00,11.00)(105.94,12.50)(105.00,14.00)
\bezier{20}(105.00,14.00)(102.94,15.50)(105.00,17.00)
\bezier{20}(105.00,17.00)(105.94,18.50)(105.00,20.00)
\put(105.00,20.00){\line(0,1){45.00}}
\bezier{20}(105.00,68.00)(102.94,69.50)(105.00,71.00)
\bezier{20}(105.00,74.00)(102.94,75.50)(105.00,77.00)
\bezier{20}(105.00,80.00)(102.94,81.50)(105.00,83.00)
\bezier{20}(105.00,86.00)(102.94,87.50)(105.00,89.00)
\bezier{20}(105.00,92.00)(102.94,93.50)(105.00,95.00)
\bezier{20}(105.00,98.00)(102.94,99.50)(105.00,101.00)
\bezier{20}(105.00,104.00)(102.94,105.50)(105.00,107.00)
\bezier{20}(105.00,65.00)(105.94,66.50)(105.00,68.00)
\bezier{20}(105.00,71.00)(105.94,72.50)(105.00,74.00)
\bezier{20}(105.00,77.00)(105.94,78.50)(105.00,80.00)
\bezier{20}(105.00,83.00)(105.94,84.50)(105.00,86.00)
\bezier{20}(105.00,89.00)(105.94,90.50)(105.00,92.00)
\bezier{20}(105.00,95.00)(105.94,96.50)(105.00,98.00)
\bezier{20}(105.00,101.00)(105.94,102.50)(105.00,104.00)
\bezier{20}(105.00,107.00)(105.94,108.50)(105.00,110.00)
\put(105.00,110.00){\line(0,1){45.00}}
\bezier{20}(105.00,158.00)(102.94,159.50)(105.00,161.00)
\bezier{20}(105.00,164.00)(102.94,165.50)(105.00,167.00)
\bezier{20}(105.00,155.00)(105.94,156.50)(105.00,158.00)
\bezier{20}(105.00,161.00)(105.94,162.50)(105.00,164.00)
\thinlines
\put(106.00,43.00){\makebox(0,0)[lc]{$R=\infty$}}
\put(106.00,87.00){\makebox(0,0)[lc]{$R=0$}}
\put(106.00,132.00){\makebox(0,0)[lc]{$R=\infty$}}
\put(59.00,132.00){\makebox(0,0)[rc]{$R=\infty$}}
\put(59.00,87.00){\makebox(0,0)[rc]{$R=0$}}
\put(59.00,43.00){\makebox(0,0)[rc]{$R=\infty$}}
\end{picture}

\vglue 1 cm
\caption{The Penrose diagram for $M>\Mcrit$ if $Q\ne0$,
and for $\Mcrit< M < 0$ if $Q=0$ and $k=-1$. There is both an outer
horizon and an inner horizon.}
\label{fig:q-penmore}
\end{figure}
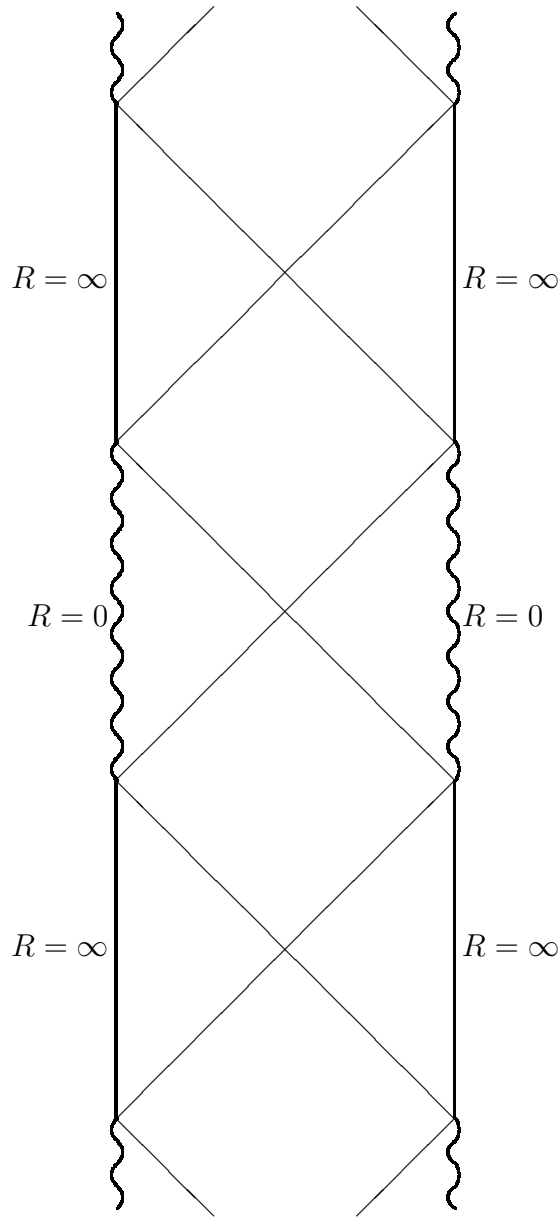

Suppose first that $Q\ne0$. For $M<\Mcrit$, $F$ has no zeros. For
$M=\Mcrit$, $F$ has a degenerate zero, and for $M>\Mcrit$, $F$ has two
distinct nondegenerate zeros. The Penrose diagrams of the analytic
extensions are shown in Figures
\ref{fig:q-penless}--\ref{fig:q-penmore}.\footnote{These statements hold
without change also for $k=1$, in which case we obtain the well known
\rnads\ spacetimes \cite{lake1,btz-cont,lou-win}.}

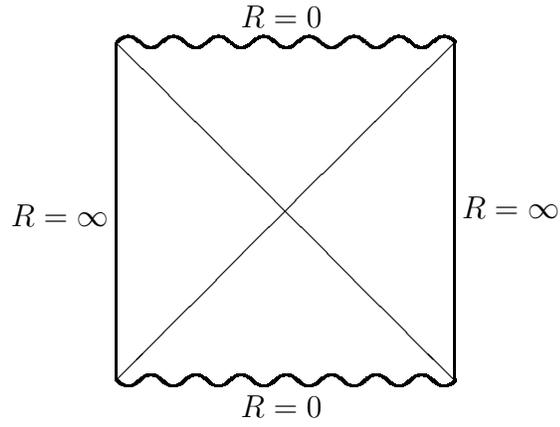
\begin{figure}
\vglue 1 cm

\unitlength 1.00mm
\linethickness{0.4pt}
\begin{picture}(106.00,57.00)
\put(105.00,55.00){\line(-1,-1){45.00}}
\put(60.00,55.00){\line(1,-1){45.00}}
\thicklines
\put(105.00,10.00){\line(0,1){45.00}}
\put(60.00,10.00){\line(0,1){45.00}}
\bezier{20}(105.00,55.00)(103.50,56.50)(102.00,55.00)
\bezier{20}(102.00,55.00)(100.50,53.50)(99.00,55.00)
\bezier{20}(99.00,55.00)(97.50,56.50)(96.00,55.00)
\bezier{20}(93.00,55.00)(91.50,56.50)(90.00,55.00)
\bezier{20}(87.00,55.00)(85.50,56.50)(84.00,55.00)
\bezier{20}(81.00,55.00)(79.50,56.50)(78.00,55.00)
\bezier{20}(75.00,55.00)(73.50,56.50)(72.00,55.00)
\bezier{20}(69.00,55.00)(67.50,56.50)(66.00,55.00)
\bezier{20}(63.00,55.00)(61.50,56.50)(60.00,55.00)
\bezier{20}(96.00,55.00)(94.50,53.50)(93.00,55.00)
\bezier{20}(90.00,55.00)(88.50,53.50)(87.00,55.00)
\bezier{20}(84.00,55.00)(82.50,53.50)(81.00,55.00)
\bezier{20}(78.00,55.00)(76.50,53.50)(75.00,55.00)
\bezier{20}(72.00,55.00)(70.50,53.50)(69.00,55.00)
\bezier{20}(66.00,55.00)(64.50,53.50)(63.00,55.00)
\bezier{20}(105.00,10.00)(103.50,8.50)(102.00,10.00)
\bezier{20}(102.00,10.00)(100.50,11.50)(99.00,10.00)
\bezier{20}(99.00,10.00)(97.50,8.50)(96.00,10.00)
\bezier{20}(93.00,10.00)(91.50,8.50)(90.00,10.00)
\bezier{20}(87.00,10.00)(85.50,8.50)(84.00,10.00)
\bezier{20}(81.00,10.00)(79.50,8.50)(78.00,10.00)
\bezier{20}(75.00,10.00)(73.50,8.50)(72.00,10.00)
\bezier{20}(69.00,10.00)(67.50,8.50)(66.00,10.00)
\bezier{20}(63.00,10.00)(61.50,8.50)(60.00,10.00)
\bezier{20}(96.00,10.00)(94.50,11.50)(93.00,10.00)
\bezier{20}(90.00,10.00)(88.50,11.50)(87.00,10.00)
\bezier{20}(84.00,10.00)(82.50,11.50)(81.00,10.00)
\bezier{20}(78.00,10.00)(76.50,11.50)(75.00,10.00)
\bezier{20}(72.00,10.00)(70.50,11.50)(69.00,10.00)
\bezier{20}(66.00,10.00)(64.50,11.50)(63.00,10.00)
\thinlines
\put(106.00,33.00){\makebox(0,0)[lc]{$R=\infty$}}
\put(59.00,32.00){\makebox(0,0)[rc]{$R=\infty$}}
\put(82.00,57.00){\makebox(0,0)[cb]{$R=0$}}
\put(82.00,8.00){\makebox(0,0)[ct]{$R=0$}}
\end{picture}

\vglue 1 cm
\caption{The Penrose diagram for $M>0$ if $Q=0$, and for
$M=0$ if $Q=0$ and $k=-1$. If $\Sigma_{-1}$ is simply connected, the
singularity in the latter case is a coordinate one.}
\label{fig:square}
\end{figure}

Suppose next that $Q=0$ and $k=-1$. We now have 
$\Mcrit = - \ell/(3\sqrt{3})$. 
For $M<\Mcrit$, $F$ has no zeros. For
$M=\Mcrit$, $F$ has a degenerate zero, and for
$\Mcrit<M < 0$, $F$ has two distinct nondegenerate zeros. The
Penrose diagrams of the analytic extensions are again as in Figures
\ref{fig:q-penless}--\ref{fig:q-penmore}. For $M\ge0$, $F$ has just
one nondegenerate zero. When $M>0$, $R=0$ is a curvature singularity, and
the Penrose diagram is shown in Figure~\ref{fig:square}. When $M=0$,
$R=0$ is not a curvature singularity, as discussed above; however,
provided $\Sigma_{-1}$ is not simply connected, we regard $R=0$ as a
topological singularity, and the Penrose diagram is again as in
Figure~\ref{fig:square}. When $M=0$ and $\Sigma_{-1}$ is simply connected,
the Penrose diagram in our coordinates is as in Figure~\ref{fig:square},
but the singularity at $R=0$ is only a coordinate one.

\begin{figure}
\vglue 1 cm

\unitlength 1.00mm
\linethickness{0.4pt}
\begin{picture}(106.00,55.95)
\thicklines
\bezier{24}(105.00,55.00)(102.11,55.95)(103.00,53.00)
\bezier{28}(103.00,53.00)(104.25,49.76)(101.00,51.00)
\bezier{24}(101.00,51.00)(98.11,51.95)(99.00,49.00)
\bezier{24}(97.00,47.00)(94.11,47.95)(95.00,45.00)
\bezier{24}(93.00,43.00)(90.11,43.95)(91.00,41.00)
\bezier{24}(89.00,39.00)(86.11,39.95)(87.00,37.00)
\bezier{24}(85.00,35.00)(82.11,35.95)(83.00,33.00)
\bezier{28}(99.00,49.00)(100.25,45.76)(97.00,47.00)
\bezier{28}(95.00,45.00)(96.25,41.76)(93.00,43.00)
\bezier{28}(91.00,41.00)(92.25,37.76)(89.00,39.00)
\bezier{28}(87.00,37.00)(88.25,33.76)(85.00,35.00)
\bezier{24}(105.00,9.00)(102.11,8.05)(103.00,11.00)
\bezier{28}(103.00,11.00)(104.25,14.24)(101.00,13.00)
\bezier{24}(101.00,13.00)(98.11,12.05)(99.00,15.00)
\bezier{24}(97.00,17.00)(94.11,16.05)(95.00,19.00)
\bezier{24}(93.00,21.00)(90.11,20.05)(91.00,23.00)
\bezier{24}(89.00,25.00)(86.11,24.05)(87.00,27.00)
\bezier{24}(85.00,29.00)(82.11,28.05)(83.00,31.00)
\bezier{28}(99.00,15.00)(100.25,18.24)(97.00,17.00)
\bezier{28}(95.00,19.00)(96.25,22.24)(93.00,21.00)
\bezier{28}(91.00,23.00)(92.25,26.24)(89.00,25.00)
\bezier{28}(87.00,27.00)(88.25,30.24)(85.00,29.00)
\bezier{12}(83.00,33.00)(83,32)(82.00,32.00)
\bezier{12}(82.00,32.00)(83,32)(83.00,31.00)
\put(105.00,9.00){\line(0,1){46.00}}
\thinlines
\put(82.00,32.00){\circle*{1.00}}
\put(106.00,33.00){\makebox(0,0)[lc]{$R=\infty$}}
\put(82.00,48.00){\makebox(0,0)[lc]{$R=0$}}
\put(82.00,18.00){\makebox(0,0)[lc]{$R=0$}}
\end{picture}

\vglue 1 cm
\caption{The Penrose diagram for $M=0$, if $Q=0$ and $k=0$.
If $\Sigma_0$ is simply connected, the singularity is a coordinate one.}
\label{fig:triangle}
\end{figure}
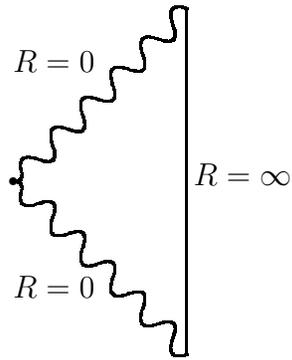

Suppose finally that $Q=0$ and $k=0$. We now have $\Mcrit = 0$. For
$M<0$, $F$ has no zeros, and for $M>0$, $F$ has a single nondegenerate
zero. The Penrose diagrams of the analytic extensions are respectively as
in Figures \ref{fig:q-penless} and~\ref{fig:square}.
In the special case $M=0$, $F$ has no zeros, and the space is locally \ads.
The Penrose diagram is shown in Figure~\ref{fig:triangle}. The status of
$R=0$ is then as above: if $\Sigma_0$ is multiply connected, we regard
$R=0$ as a topological singularity, whereas if $\Sigma_0$ is simply
connected, $R=0$ is just a coordinate singularity.

We have therefore obtained Penrose diagrams that faithfully depict the
causal structure of the spacetimes, with the sole exception of $M=0=Q$
and $\Sigma_k$ simply connected. With this exception, we see that all
the spacetimes in which $F$ has a nondegenerate zero can be interpreted
as black holes. The connected components of the infinities displayed in
the Penrose diagrams are genuine future null infinities, and the
boundaries of their causal pasts are black hole horizons. When a second
zero of $F$ exists, it can be interpreted as an inner horizon, as in the
\rnads\ spacetime \cite{lake1,btz-cont}. The topology of the horizon
bifurcation two-manifold is that of~$\Sigma_k$, and thus different
from~$S^2$. The theorems about spherical horizon topology
\cite{topocen,galloway,chrus-wald,jacobson-venka} do not apply because the
negative cosmological constant can be interpreted as a negative vacuum
energy density.\footnote{For discussions of
$k=1$ with the $\RPtwo$ horizon topology but without a cosmological
constant, see Refs.\ \cite{topocen,chamb-gibb}.}

In the spacetimes in which $F$ has a degenerate zero, it is seen from
figure \ref{fig:q-peneq} that the future null infinity consists of a
single connected component, and the past of this infinity is all
of the spacetime. The Killing horizons in these spacetimes therefore do
not have an interpretation as black hole horizons. Note that this differs
from the extreme  Reissner-Nordstr\"om solutions \cite{haw-ell}, in which
the future null infinity is not connected, and the past of each connected
component has a boundary along a Killing horizon.

The existence criterion for a nondegenerate horizon is $M>\Mcrit$. For
$k=0$, we have
$\Mcrit \ge 0$, and black holes therefore only occur with
positive values of~$M$. For $k=-1$, however, $\Mcrit$ is negative for $|Q|<
\ell/2$, so that black holes occur even with negative values of~$M$. Note
also that when $k=-1$ and $Q=0$, the internal structure of the black hole
changes qualitatively at $M=0$: for 
$- \ell /(3\sqrt{3})<M<0$, 
we have two horizons and the singularities are timelike
(Figure~\ref{fig:q-penmore}), whereas for $M>0$, we only have one horizon
and the singularities are spacelike (Figure~\ref{fig:square}). Provided
$\Sigma_{-1}$ is not simply connected, we regard the limiting case $M=0$ as
belonging to the latter category.

Instead of the pair $(M,Q)$, it is more convenient to parametrize the
black hole spacetimes in terms of the pair $(\Rh,Q)$,
where $\Rh$ is the value of $R$ at the (outer) horizon. For given~$Q$,
$\Rh$ can take the values $\Rh > \Rcrit(Q)$, where
\begin{equation}
\Rcrit(Q) :=
{\ell\over\sqrt{6}}
\left( \sqrt{k^2 + 12 {(Q/\ell)}^2} - k \right)^{1/2}
\ \ .
\label{Rcrit}
\end{equation}
The mass is then given in terms of $Q$ and $\Rh$ as
\begin{equation}
M = {\Rh \over 2}
\left( {\Rh^2 \over \ell^2} + k  + {Q^2 \over \Rh^2} \right)
\ \ .
\label{massfunc}
\end{equation}

\section{Thermodynamics with finite boundary}
\label{sec:finite-thermo}

In this section we consider the thermodynamics of a black hole in a
finite size box.  First we calculate the local Hawking temperature for the
black hole both by using the surface gravity formula, and by identifying
the periodicity in the time coordinate in the Euclideanized metric. Then,
we put the black hole  in a box and use the Brown-York quasilocal energy
formalism to calculate what we call the thermodynamical internal energy
for the system. Upon varying this energy with respect to the extensive
variables $M$ and~$Q$, using the expression for the local Hawking
temperature, and assuming that the first law of black hole thermodynamics
holds, we identify the entropy and the  electrostatic potential for the
system. Finally, we calculate the signs of the heat capacities
$C_Q$, $C_{\Phi_B}$ and~$C_{\phi_{\rm h}}$.
We include the three cases $k=1$, $k=0$, and $k=-1$ throughout the
section.

\subsection{Local Hawking temperature}

The local Hawking temperature for a static eternal black hole can be
calculated using the Unruh effect in curved spacetime or
finding the periodicity in the time coordinate in the
Euclidean version of the black hole metric covering the outer region.
(See for example Ref.\ \cite{wald-qft}.)

In the Unruh effect one considers how an observer outside the black
hole\footnote{The Unruh effect gives rise to a Hawking temperature even
without a black hole, as long as the spacetime contains a bifurcate
Killing horizon, there exists a Killing field timelike in the outer
region, and the Hartle-Hawking vacuum exists
\cite{wald-qft}.}, following the timelike Killing flow, would experience
a quantum field that is in the Hartle-Hawking vacuum state. The
Hartle-Hawking vacuum is a globally non-singular vacuum invariant under
the Killing flow. The result is that the observer will experience a
thermal state with local temperature
\begin{mathletters}
\begin{equation}
T_H(R)=\frac{\kappa_{\rm h}}{2 \pi \sqrt{-\chi _{\alpha}\chi^{\alpha}}}
\ \ ,
\label{hawking-temp}
\end{equation}
where $\kappa_{\rm h}$ is the surface gravity evaluated at the horizon,
\begin{equation}
\kappa_{\rm h} := \left.
\sqrt{-\casehalf \nabla ^{\alpha}\chi ^{\beta}\nabla
_{\alpha}\chi _{\beta}}
\right | _{R=\Rh}
\ \ ,
\label{surface-gravity}
\end{equation}
\end{mathletters}%
and $\chi^{\alpha}$ is the Killing vector field generating the event
horizon. The interpretation of this physical process is that we have a
black hole in thermal equilibrium with its surroundings.

For the black holes (\ref{gen-metric}) considered here, the Killing field
is
\begin{equation}
\chi ^{\alpha}\frac{\partial}{\partial
x^{\alpha}}=\frac{\partial}{\partial T}
\ \ .
\label{killing}
\end{equation}
We therefore have
\begin{mathletters}
\begin{equation}
\kappa = \casehalf F'(R) \label{kappa}
\ \ ,
\end{equation}
where the prime indicates derivative with respect to~$R$, and
\begin{equation}
T_H(R)={ F'(\Rh) \over 4\pi \sqrt{F(R)}}
=
{\left(
k
- Q^2/\Rh^2
+ 3 \Rh^2/\ell^2
\right )
\over
4\pi \Rh
\sqrt{k
- 2M/R
+ Q^2/R^2
+ R^2/\ell^2 }}
\ \ .
\label{hawking-temp2}
\end{equation}
\end{mathletters}%
Note that $T_H(R)$ does not depend on the normalization of the Killing
vector field $\chi^{\alpha}$. In the limits $R\rightarrow \Rh$ and
$R\rightarrow \infty$, we have respectively $T_H(R)\rightarrow \infty$
and $T_H(R)\rightarrow 0$. Unlike in the asymptotically flat case, the
black hole therefore does
not have a finite, nonvanishing physical temperature at infinity.

It is of interest to define the renormalized temperature, denoted
by~$T_\infty$, as the product of
$T_H(R)$ and the redshift factor $\sqrt{-\chi^{\alpha}\chi _{\alpha}}$
\cite{HPads,PagePhill,pagerev}. The result is
\begin{equation}
T_\infty =
{ F'(\Rh) \over 4\pi }
=
{\left(
k
- Q^2/\Rh^2
+ 3 \Rh^2/\ell^2
\right )
\over
4\pi \Rh }
\ \ .
\label{hawking-temp-infty}
\end{equation}
Although $T_\infty$ does not appear to have a physical interpretation as
the temperature experienced by a family of observers\footnote{$T_\infty$
coincides with $T_H(R)$ at the locations where the redshift factor equals
unity. However, these locations depend on the normalization of the Killing
vector~$\chi^{\alpha}$.},
we shall see below that it emerges as the counterpart of temperature in the
infinite space limit of the first law of thermodynamics
\cite{HPads,PagePhill,pagerev,BrownCreMann}.

For given~$k$, both $T_H(R)$ and $T_\infty$ are independent of
the topology of the two-space~$\Sigma_k$. Also, both
$T_H(R)$ and $T_\infty$
vanish for the extremal solutions,
$M=\Mcrit$, as $F(R)$ then has a double root at $R=\Rh$.

These results for the Hawking temperature can also be derived by Euclidean
methods \cite{HPads}. When a nondegenerate horizon exists, regularity of
the Euclidean version of the metric (\ref{gen-metric}) at the
horizon requires the Euclidean time, $\tau := iT$, to be periodic with
period $P=4\pi/F'(\Rh)$. When a Green's function that is regular on the
Euclidean section is analytically continued to the Lorentzian section,
it retains periodicity in imaginary time. The local temperature can
then be identified as the inverse Euclidean period divided by the redshift
factor, with the result
$T_H(R)=(Pg_{00})^{-1}=F'(\Rh){\left[4\pi \sqrt{F(R)}\right]}^{-1}$.

\subsection{First law and entropy}

We wish to verify that the black holes satisfy the first law of
black hole thermodynamics (BHTD) and to identify the black hole entropy.
As $T_H(R)\rightarrow 0$ in the limit $R\rightarrow \infty$, we
first formulate the first law with a boundary at a finite value of~$R$. To
have a black hole spacetime, we assume throughout $M>M_{crit}$. 
To make the thermodynamical quantities finite, we take $\Sigma_k$ closed. As
in section~\ref{sec:spacetimes}, $V$ denotes the (dimensionless) area
of~$\Sigma_k$.

We introduce a boundary at $R=R_B$, and we regard the boundary scale factor
$R_B$ as a prescribed, finite parameter. A spacelike snapshot of the
boundary history  then has the topology of~$\Sigma_k$. The Brown-York
quasilocal energy formalism \cite{BY-quasilocal} can be readily used to
define the thermodynamical internal energy of this system on a constant $T$
hypersurface. Denoting the internal energy by~$U(R_B)$, we find
\begin{equation}
U(R_B)=-R_B^2 V \left ( \frac{\sqrt{F(R_B)}}
{4 \pi R_B} + \epsilon_0(R_B) \right )
\ \ ,
\label{int-energy}
\end{equation}
where $\epsilon_0(R)$ is an arbitrary function that arises from
the freedom of adding surface terms to the gravitational action. A
specific choice for $\epsilon_0(R)$ will not be needed for what follows.

We note in passing that one natural criterion for choosing $\epsilon_0(R)$
would be to require that $U(R_B)$ vanishes for the locally \ads\
solutions, for which $M=0=Q$. This leads to
\begin{equation}
\epsilon _0(R) = -\frac{\sqrt{k+ R^2/\ell ^2}}{4\pi R}
\ \ .
\end{equation}
For $k=1$ and $k=0$, we then have $U(R_B)\ge0$, but for $k=-1$, $U(R_B)$
does not have a definite sign. In particular, for $k=-1$, $U(R_B)<0$ when
$Q=0$ and $M<0$.

Variation of $U(R_B)$ with respect to $M$ and~$Q$ (or, equivalently, $\Rh$
and~$Q$) gives
\begin{eqnarray}
dU(R_B)&=&
{V
\over
8 \pi
\sqrt{F(R_B)}}
\left(k - {Q^2\over \Rh^2} + {3 \Rh^2 \over \ell^2}
\right)
d\Rh
+
\frac{QV}{4\pi \sqrt{F(R_B)}}\left
(\frac{1}{\Rh}-\frac{1}{R_B}\right ) dQ \nonumber \\
& &\nonumber \\
&=&T_H(R_B) \, d(\case{1}{4} V \Rh^2) +
\tilde{\Phi}(R_B) de
\ \ ,
\label{first-law}
\end{eqnarray}
where we have used the Hawking temperature
(\ref{hawking-temp2}) and defined
\begin{mathletters}
\begin{eqnarray}
e&:=& (V/4\pi)Q
\label{e-def}
\ \ ,
\\
\tilde{\Phi}(R_B) &:=& \frac{Q}{\sqrt{F(R_B)}}\left
(\frac{1}{\Rh}-\frac{1}{R_B}\right )
\ \ .
\label{Phitilde}
\end{eqnarray}
\end{mathletters}%
As mentioned in section~\ref{sec:spacetimes}, $e$ is the ADM
charge. Comparing (\ref{Phitilde}) to (\ref{A}) shows that
$\tilde{\Phi}(R_B)$ is equal to the electrostatic potential difference
between the horizon and the boundary, with the electromagnetic gauge
chosen as in~(\ref{A}), and with respect to a time  coordinate that agrees
with the proper time of a static observer at the boundary. We can think of
$\tilde{\Phi}(R_B)$ as the electrostatic potential difference between the
horizon and the boundary, appropriately redshifted to the boundary.

Comparing (\ref{first-law}) with the desired form of the first law of
BHTD,
\begin{equation}
dU=TdS + {\tilde\Phi} de
\ \ ,
\end{equation}
we identify the entropy of the black hole as
\begin{equation}
S=\case{1}{4} V \Rh^2 = \case{1}{4} A_{\rm h}
\ \ ,
\label{entropy}
\end{equation}
where $A_{\rm h}$ is the area of the event horizon.
This area law holds for all the closed horizon topologies that occur with
our black hole spacetimes.
In the special case $k=1$ and $\Sigma_1=S^2$,
we recover the Bekenstein-Hawking area law.

It would be possible to vary $U(R_B)$ also with respect to~$R_B$. The
first law (\ref{first-law}) would then contain the additional term
$-p_B d R_B$, where $p_B$ is the surface pressure,
thermodynamically conjugate to~$R_B$ \cite{york1,B+}. The expression for
$p_B$ would, however, depend on the choice of the term $\epsilon_0(R)$
in~(\ref{int-energy}).

In the limit $R_B\to\infty$, both sides of (\ref{first-law})
vanish. Nevertheless,
multiplying first both sides by $\sqrt{F(R_B)}$ and then taking
the limit $R_B\to\infty$, we recover the finite equation
\begin{equation}
d\left(VM \over 4\pi\right)
=
T_\infty \, d(\case{1}{4} V \Rh^2) +
\phi de
\ \ ,
\label{infty-first-law}
\end{equation}
where
\begin{equation}
\phi := \frac{Q}{\Rh}
\ \ .
\label{phi-infty}
\end{equation}
{}From section \ref{sec:spacetimes} we recall that $(V/4\pi)M$ is the ADM
energy at infinity and $\phi$ is the electrostatic potential
difference between the horizon and infinity, both with respect to the
Killing time coordinate of the metric~(\ref{gen-metric}). We can therefore
identify equation (\ref{infty-first-law}) as the first law of
BHTD in the absence of a boundary. If $T_\infty$ is postulated to have an
interpretation as a temperature, we obtain for the entropy the area law
(\ref{entropy}) \cite{HPads}. Conversely, if the area law (\ref{entropy})
for the entropy is postulated to hold, $T_\infty$ emerges as a
temperature \cite{BrownCreMann}.
We re-emphasize, however, that $T_\infty$ is not the physical temperature
measured by an observer at infinity.

Note that the first laws (\ref{first-law}) and (\ref{infty-first-law})
only determine the entropy up to an additive constant.
In the identification (\ref{entropy}) we have chosen this constant so that
the entropy is 
equal to one quarter of the area. 
One could, however, add to
(\ref{entropy}) an arbitrary function of any quantities that our variations
treat as fixed. In particular, one could add an arbitrary function of the
cosmological constant and the topology of~$\Sigma_k$. 

Finally, we note that the above thermodynamical discussion has regarded $V$
as fixed. It does not appear possible to relax this assumption in a way
that would promote $V$ into an independent thermodynamical variable. For
$k=1$ and $k=-1$, $V$ only takes discrete values, and continuous
variations in $V$ are not possible. For $k=0$, the possible values of $V$
form a continuum; however, changes in $V$ can then be absorbed into
redefinitions of $R$, $M$, and~$Q$.

\subsection{Thermodynamical stability}
\label{subsec:finite-thermo-stability}

We now turn to the thermodynamical stability of the black holes. In this
section we consider a black hole in a box with a prescribed, finite value
of~$R_B$. The limit $R_B\to\infty$ will be addressed in
section~\ref{sec:infinite-thermo}.

The response function whose sign determines the thermodynamical
stability is the heat capacity (see for example Ref.\ \cite{reichl})
\begin{equation}
C_X = T
\left( \frac{\partial S}{\partial T}\right) _{X}
\ \ ,
\label{cm}
\end{equation}
where $S$ is the entropy, $T$ is the temperature, and $X$ indicates the
quantities that are held fixed. With a finite boundary, the relevant
temperature is the local Hawking temperature~(\ref{hawking-temp2}), and
the entropy is given by the area law~(\ref{entropy}). For the fixed
quantity~$X$, we consider three choices: the ADM charge $e$~(\ref{e-def})
(or, equivalently, the parameter~$Q$), the redshifted electrostatic
potential difference $\tilde{\Phi}(R_B)$ [Eq.~(\ref{Phitilde})]
between the horizon and the boundary \cite{BBWY}, and the redshifted
electrostatic potential difference between the
boundary and infinity, given by
\begin{equation}
\Phi_B := {Q \over R_B \sqrt{F(R_B)} }
\ \ .
\label{PhiB-def}
\end{equation}
We write $\tilde{\Phi}(R_B):=\tilde{\Phi}_B$.

The technical details of analyzing the three heat capacities $C_Q$,
$C_{\tilde{\Phi}_B}$ and $C_{\Phi _B}$ are given in the appendix. When
$R_B$ is so large that the box-dependent features of the heat capacities
become negligible, we find that these heat capacities are positive definite
for $k=0$ and $k=-1$, but indefinite for $k=1$. In this sense, the black
holes with $k=0$ and $k=-1$ have a wider range of thermodynamical stability
than the conventional black holes with $k=1$. However, as discussed in the
appendix, there exist choices for the fixed quantity $X$ that would render
also the black holes with $k=0$ and $k=-1$ thermodynamically
unstable.

For the conventional black holes with $k=1$, the heat capacities $C_Q$,
$C_{\tilde{\Phi}_B}$ and $C_{\Phi _B}$ diverge at the places where they
change sign in the $(M,Q)$ parameter space. In the asymptotically flat
context, this phenomenon was discussed by Davies \cite{davies1,davies2}.

\section{Infinite space thermodynamical ensembles}
\label{sec:infinite-thermo}

In this section we consider thermodynamics in the limit where
the boundary is pushed to infinity. For $k=1$ and $\Sigma_1=S^2$, this
problem was analyzed in Refs.\ \cite{BrownCreMann,lou-win}. We take here
$k=0$ or $k=-1$, and assume throughout that $\Sigma_k$ is closed.

As discussed in section~\ref{sec:finite-thermo}, both the local Hawking
temperature $T_H(R_B)$ (\ref{hawking-temp2}) and the redshifted
electrostatic  potential difference $\tilde{\Phi}(R_B)$
(\ref{Phitilde}) vanish in the limit $R_B\to\infty$. Relying on the
infinite space form (\ref{infty-first-law}) of the first law, we adopt the
viewpoint that the appropriate counterparts of $T_H(R_B)$
and $\tilde{\Phi}(R_B)$ are, respectively, the renormalized temperature
$T_\infty$ (\ref{hawking-temp2}) and the (unredshifted) Killing time
electrostatic potential difference $\phi$~(\ref{phi-infty}). We
write $\binfty = T_\infty^{-1}$.

It would be straightforward to proceed as in section \ref{sec:finite-thermo}
and show that the heat capacities at fixed $e$ (\ref{e-def}) and
$\phi$ (\ref{phi-infty}) are both positive definite. However, we wish to go
further and construct full quantum thermodynamical equilibrium ensembles
that fix, in addition to $\binfty$, either $e$ or~$\phi$. Following the
terminology of Refs.\
\cite{lou-win,WYprl,whitingCQG,BBWY,LW2,BLPP,LouSiWi}, we refer to the
ensemble that fixes $\binfty$ and $e$ as the canonical ensemble, and 
to the ensemble that fixes $\binfty$ and $\phi$ as the grand canonical
ensemble.

One way to approach this problem would be within the Euclidean
path-integral formalism, performing a Hamiltonian reduction of the action
as in Refs.\ \cite{WYprl,whitingCQG,BBWY}. Another way would be to perform
a Hamiltonian reduction in the Lorentzian theory, and then take the trace
of an analytically continued evolution operator under suitably chosen
boundary conditions as in Refs.\ \cite{lou-win,LW2,BLPP,LouSiWi}. The
boundary conditions in the two approaches are identical by construction,
and one may argue that the only difference between
the two approaches is in the order of quantization and Euclideanization.
For our spacetimes, the appropriate boundary conditions and boundary
terms are easily found by adapting to our symmetries the Lorentzian
Hamiltonian analysis of the $k=1$ case in Ref.\ \cite{lou-win}. Adapting to
our symmetries the details of the Lorentzian Hamiltonian reduction of
Ref.\ \cite{lou-win} would require more work, and we have not pursued this
in detail; instead, we appeal to the Euclidean reduction formalism
\cite{WYprl,whitingCQG,BBWY} to argue that only the boundary terms survive
after the reduction. This yields the reduced actions through steps that
follow the cited references so closely that we shall not repeat the
details of the analysis here. Instead, we just state the results for
the reduced Euclidean actions, and proceed to the thermodynamical
analysis.

\subsection{Grand canonical ensemble}
\label{subsec:grand-can}

The reduced Euclidean action with fixed $\binfty$ and $\phi$ is given by
\begin{equation}
I^*_{\rm gc} (\Rnought,\bq) :=
{V \over 4\pi} \left[
\binfty (\bm - \bq \phi) - \pi \Rnought^2
\right]
\ \ ,
\end{equation}
where
\begin{equation}
\bm :=
\casehalf \Rnought
\left( \Rnought^2 \ell^{-2} + k + \bq^2 \Rnought^{-2} \right)
\ \ .
\label{massfunc2}
\end{equation}
The variables in $I^*_{\rm gc}$ are $\Rnought$ and~$\bq$, and their domain
is specified by the inequalities
\begin{mathletters}
\label{int-domain}
\begin{eqnarray}
&&\Rnought > \sqrt{-k/3} \, \ell
\ \ ,
\\
&&\bq^2 < \Rnought^2 \left( k + 3 \Rnought^2 \ell^{-2} \right)
\ \ .
\label{ineqb}
\end{eqnarray}
\end{mathletters}%
The reduction has eliminated the constraints, but it has not used the full
Einstein equations. For generic values of $\Rnought$ and~$\bq$, $I^*_{\rm
gc} (\Rnought,\bq)$ is therefore not equal to the Euclidean action of any
of the classical black holes of section~\ref{sec:spacetimes}. However,
$I^*_{\rm gc} (\Rnought,\bq)$ is the Euclidean action of a spacetime with
the same topological and asymptotic properties. In particular, $\Rnought$
is the value of the ``scale factor" associated with $\Sigma_k$ at the
horizon, and the ADM charge at infinity is $(V/4\pi)\bq$.

$I^*_{\rm gc}$ has precisely one stationary point, at
\begin{mathletters}
\label{gc-critical}
\begin{eqnarray}
\Rnought = \Rnought^+
&:=& {2\pi \ell^2 \over 3 \binfty}
\left[ 1 + \sqrt{1 +  {3\binfty^2 (\phi^2 - k) \over 4 \pi^2 \ell^2}}
\right]
\ \ ,
\\
\bq = \bq^+ &:=& \phi \Rnought^+
\ \ ,
\end{eqnarray}
\end{mathletters}%
and this stationary point is the global minimum. It is straightforward
to verify that this stationary point is the black hole spacetime of
section \ref{sec:spacetimes} with the specified values of $\binfty$
and~$\phi$. $\Rnought^+$ is equal to the value of $\Rh$ in this spacetime,
$\bq^+$ is equal to the value of~$Q$, and $\bm^+:=\bm(\Rnought^+,\bq^+)$
is equal the value of~$M$.

The grand partition function of the thermodynamical grand canonical
ensemble is obtained  as the integral
\begin{equation}
{\cal Z} (\binfty,\phi)
= \int_{{\cal A}} {\tilde \mu}
\, d\Rnought d\bq \,
\exp\left(  -I^*_{\rm gc} \right)
\ \ ,
\label{cZ}
\end{equation}
where the integration domain ${\cal A}$ is given
by~(\ref{int-domain}). The weight factor~${\tilde \mu}$, which depends on
the details of quantization
\cite{lou-win,WYprl,whitingCQG,BBWY,LW2,LW1}, is assumed to be positive and
slowly varying. The qualitative properties of the ensemble are then
determined by the exponential factor in~(\ref{cZ}).

The integral in (\ref{cZ}) is convergent, and when the stationary point
approximation is good, the dominant contribution comes from the global
minimum at the stationary point~(\ref{gc-critical}). Denoting by $\langle
E \rangle$ and $\langle e \rangle$ the thermal expectation values of
respectively the energy and the charge, we have
\begin{mathletters}
\begin{eqnarray}
\langle E \rangle &=&
\left( - {\partial \over \partial \binfty}
+ \binfty^{-1} \phi {\partial \over \partial \phi} \right)
(\ln {\cal Z})
\approx
(V/4\pi) \bm^+
\ \ ,
\\
\langle e \rangle &=&
\binfty^{-1} {\partial (\ln {\cal Z}) \over \partial \phi}
\approx
(V/4\pi) \bq^+
\ \ .
\end{eqnarray}
\end{mathletters}%
It follows from the construction of the grand canonical ensemble that the
constant $\phi$ heat capacity, $C_\phi = \binfty^2 \biglb(\partial^2
(\ln {\cal Z}) / \partial\binfty^2\bigrb)$, is positive, and
also that $(\partial\langle e \rangle /
\partial \phi)$ is positive: when the stationary point approximation is
good, these statements can be easily verified observing that $\partial
\Rnought^+ / \partial\binfty<0$ and $\partial \bq^+ / \partial
\phi>0$. The system is therefore stable under thermal fluctuations in both
the energy and the charge.

When the stationary point dominates, we obtain for the entropy
\begin{equation}
S = \left( 1 - \binfty {\partial \over \partial \binfty} \right)
(\ln {\cal Z})
\approx
\case{1}{4} V {(\Rnought^+)}^2
=
\case{1}{4} A_{\rm h}
\ \ .
\end{equation}
This agrees with the area law~(\ref{entropy}).

\subsection{Canonical ensemble}
\label{subsec:can}

In the canonical ensemble, we wish to fix $\binfty$ and the ADM charge~$e$.
The reduced Euclidean action with these fixed
quantities is
\begin{equation}
I^*_{\rm c} (\Rnought) :=
{V \over 4\pi} \left(
\binfty \bm - \pi \Rnought^2
\right)
\ \ ,
\end{equation}
where $\bm$ is given by (\ref{massfunc2}) with $\bq=(4\pi/V)e$. The
only variable in $I^*_{\rm c}$ is~$\Rnought$, and its domain is
$\Rnought>\Rcrit(\bq)$, where the function $\Rcrit$ was defined
in~(\ref{Rcrit}). Again, the reduction has eliminated the constraints but
not used the full Einstein equations, and for generic values
of~$\Rnought$, $I^*_{\rm c} (\Rnought)$ is not equal to the Euclidean
action of any of the black holes of section~\ref{sec:spacetimes}. Instead,
$I^*_{\rm c} (\Rnought)$ is the Euclidean action of a spacetime with
the same topological and asymptotic properties, and $\Rnought$
is the value of the ``scale factor" of $\Sigma_k$ at the horizon of this
spacetime.

$I^*_{\rm c}$ has precisely one stationary point, at the unique
root of the equation
\begin{equation}
{3\Rnought^4 \over \ell^2}-
{4\pi \Rnought^3 \over \binfty} +k \Rnought^2 -  \bq^2
= 0
\end{equation}
in the domain $\Rnought>\Rcrit(\bq)$. This stationary point is the global
minimum of~$I^*_{\rm c}$. It is straightforward to verify that
this stationary point is the black hole spacetime of section
\ref{sec:spacetimes} with the specified values of $\binfty$ and~$e$.
$\Rnought^+$ is equal to the value of $\Rh$ in this spacetime, $Q=\bq
= (4\pi/V)e$, and $\bm(\Rnought^+)$ is equal the value of~$M$.

The partition function of the thermodynamical canonical ensemble reads
\begin{equation}
Z (\binfty,e) = \int\limits_{\Rcrit(\bq)}^\infty
{\tilde{\tilde\mu}}
\, d\Rnought \,
\exp\left(  -I^*_{\rm c} \right)
\ \ ,
\label{Z}
\end{equation}
where we again assume the weight factor ${\tilde{\tilde\mu}}$ to be
positive and slowly varying compared with the exponential. The integral is
convergent, and the positivity of the constant $e$ heat capacity,
$C_e = \binfty^2 \biglb(\partial^2 (\ln Z) / \partial\binfty^2\bigrb)$, is
guaranteed by construction. When the stationary point approximation is
good, the dominant contribution comes from the unique stationary point. For
the thermal expectation values of the energy and the electric potential, we
find
\begin{mathletters}
\begin{eqnarray}
\langle E \rangle &=&
 - {\partial (\ln Z) \over \partial \binfty}
\approx (V/4\pi)\bm
\ \ ,
\label{can-E-expect}
\\
\langle \phi \rangle &=&
- \binfty^{-1} {\partial (\ln Z) \over \partial e}
\approx
{\bq \over
\Rnought}
\ \ ,
\end{eqnarray}
\end{mathletters}%
which are related to the parameters of the
dominating classical solution in the expected way. When the approximation
in (\ref{can-E-expect}) for $\langle E \rangle$ holds, the positivity of
$C_e$ can be verified observing that at the critical point
$\partial\bm/\partial\binfty < 0$. For the entropy we
again recover the
area law~(\ref{entropy}),
\begin{equation}
S = \left( 1 - \binfty {\partial \over \partial \binfty} \right)
(\ln Z)
\approx
\case{1}{4} A_{\rm h}
\ \ .
\end{equation}

\section{Discussion}
\label{sec:discussion}

In this paper we have discussed the thermodynamics of asymptotically \ads\
black holes in which the round two-sphere of the \rnads\ spacetimes has been
replaced by a two-dimensional space $\Sigma$ of constant negative or
vanishing curvature. The local properties of these spacetimes are well
known \cite{exact-book}. The main new feature for black hole
interpretation is that the topology of the horizon is not spherical but
that of~$\Sigma$. This allows a toroidal horizon when $\Sigma$ is flat, and
a horizon with the topology of any closed higher genus Riemann surface when
$\Sigma$ has negative curvature. More possibilities arise if $\Sigma$ is
not closed.

All the spacetimes have one or more asymptotically \ads\ infinities, and one
can use the asymptotic Killing time translations to define ADM mass and
charge. These quantities are finite whenever $\Sigma$ is closed. If the
additive constant in the ADM mass is chosen so that the mass vanishes for
the solutions that are locally \ads, black holes with flat $\Sigma$ have
positive ADM mass, but when $\Sigma$ has negative curvature, there are
black hole solutions with either sign of the ADM mass.

The thermodynamical analysis was carried out via a straightforward
generalization of the techniques previously applied to the \rnads\
spacetimes. The local Hawking temperature was found from the Unruh
effect, or from the periodicity of Euclidean time. Taking $\Sigma$
closed, we introduced a boundary with the topology of~$\Sigma$, and we
interpreted the Brown-York type quasilocal energy at the boundary as the
internal thermodynamical energy. The first law of black hole thermodynamics
then led to the conclusion that the entropy is one quarter of the horizon
area. This result extends the Bekenstein-Hawking area law to our toroidal
and higher genus horizons.

Examination of heat capacities with fixed ADM charge, or with
fixed appropriate electrostatic potentials, showed that our black holes
are thermodynamically more stable than the \rnads\ black hole. In
particular, in the limit of a large box, our black holes are always
thermodynamically stable under these boundary conditions. With the boundary
pushed fully to infinity, we constructed quantum equilibrium
ensembles that fixed a renormalized temperature and either the ADM charge
or the electrostatic potential. We found that these ensembles are well
defined, and always dominated by a unique black hole solution. This provides
another piece of evidence for thermodynamical stability of our black holes.

All our black hole spacetimes belong to the family~(\ref{gen-metric}). This
family arises as the generic solution family from a Birkhoff's
theorem that assumes the spacetime to admit the local isometry group
$\etwo$ (leading to flat~$\Sigma$), $\soctwoone$ (leading to negatively
curved~$\Sigma$), or
$\sothree$ (leading to the \rnads\ solutions), with two-dimensional
spacelike orbits \cite{exact-book}.
This suggests seeking a black hole interpretation also for spacetimes that
have the same local isometries but do not fall within the
family~(\ref{gen-metric}). The most promising candidate would seem to be
the Nariai-Bertotti-Robinson family \cite{exact-book}. 
In this family, the spacetime has the product form $M_L\times M_E$, where
$M_L$ ($M_E$, respectively) is a two-dimensional Riemannian manifold of
signature
$(-+)$
$\bigl((++)\bigr)$ and constant Gaussian curvature $K_L$~($K_E$). The
curvatures satisfy
\begin{mathletters}
\begin{eqnarray}
&&K_E + K_L = 2 \Lambda
\ \ ,
\\
&&K_E - K_L \ge 0
\ \ .
\end{eqnarray}
\end{mathletters}%
The electromagnetic two-form with a vanishing magnetic field is
\begin{equation}
\bbox{F} = \pm \sqrt{\casehalf(K_E - K_L)}
\,
\bbox{\omega}_L
\ \ ,
\end{equation}
where $\bbox{\omega}_L$ is the volume two-form on~$M_L$, and the case of a
nonvanishing magnetic field is obtained via the electromagnetic
duality rotation.\footnote{The special case $\bbox{F}=0$ yields flat
spacetime for $\Lambda=0$, the Nariai solution \cite{nariai} for
$\Lambda>0$, and a negative curvature analogue of the Nariai solution for
$\Lambda<0$. The special case $\Lambda=0$ yields the Bertotti-Robinson
solution \cite{levicivita,bertotti,robinson}
for $\bbox{F}\ne0$ and flat spacetime for
$\bbox{F}=0$.} The local symmetries of the spacetime are clear from the
construction. To create a black hole in analogy with the BTZ construction
\cite{btz-hole,carlip-rev,ABHP,banados4}, one would now like to take the
quotient with respect to a suitable discrete isometry group. The crucial
question is whether  satisfactory discrete isometries exists.

{\em Note added\/}.
After the present work was completed, Ref.\ \cite{vanzo} was posted. The
results therein overlap with ours for $Q=0$ and in the absence of a finite
boundary. The main difference is that the subtraction procedure of 
Ref.\ \cite{vanzo} to make the Euclidean action finite generates for $k=-1$
a horizon contribution that is not present in our Hamiltonian subtraction
procedure in section~\ref{sec:infinite-thermo}. As a result, the
entropy obtained  in Ref.\ \cite{vanzo} for $k=-1$ differs from
(\ref{entropy}) by an additive constant, such that the values of
the entropy span the whole positive real axis. Also, the additive constant
in the ADM energy in Ref.\ \cite{vanzo} is chosen so that the ADM energy
takes all positive values both for $k=0$ and $k=-1$.

\acknowledgments
We would like to thank Ingemar Bengtsson, S\"oren Holst, Ted Jacobson, and
Bernard Whiting for discussions. This work was supported in part by NSF
grant PHY-94-21849.

\appendix
\section*{Heat capacities in a box}
\label{app: spec-heat}

In this appendix we calculate the heat capacities~$C_Q$,
$C_{\tilde{\Phi}_B}$, and~$C_{\Phi _B}$, defined in
section~\ref{sec:finite-thermo}.
We consider both $k=1$, $k=0$, and $k=-1$.

As explained in section~\ref{sec:finite-thermo}, we consider a black hole in
a box with a prescribed, finite boundary scale factor~$R_B$. The potential
$\tilde{\Phi}_B:=\tilde{\Phi}(R_B)$ is defined by~(\ref{Phitilde}), and
it equals the electrostatic potential difference between the boundary and
the horizon, with respect to a time coordinate normalized to a static
observer's proper time at the boundary. Similarly, the potential
$\Phi_B$ was defined by~(\ref{PhiB-def}), and it equals the electrostatic
potential difference between the boundary and the infinity, with respect to
a time coordinate normalized to a static observer's proper time at the
boundary.

The heat capacity $C_X$ at constant value of the
thermodynamical variable $X$ is defined by~(\ref{cm}),
where $S$ is the entropy and $T$ the temperature. For us, $S$~and
$T=T_H(R_B)$ are given respectively by (\ref{entropy})
and~(\ref{hawking-temp2}).

It is useful to regard $S$ and $T_H$ as functions of the two
independent variables $\Rh$ and $Q^2$. $M$ becomes then a dependent
variable, determined by~(\ref{massfunc}). {}From~(\ref{cm}), we obtain
\begin{equation}
C_X =
\casehalf V\Rh T_H
\left[
{T_{\Rh} +
T_{Q^2}
\left ( \frac{d Q^2}{d\Rh}\right )_X}
\right]^{-1}
\ \ ,
\label{spec-heat2}
\end{equation}
where
\begin{mathletters}
\label{Trq}
\begin{eqnarray}
T_{\Rh}&:=&\frac{\partial T_H}{\partial \Rh}
\nonumber
\\
&=&
\frac{1 }{4 \pi
\sqrt{F(R)}}\left[ \frac{1}{\Rh^2}\left ( -k+\frac{3 \Rh^2}{\ell ^2} +
\frac{3Q^2}{\Rh^2}\right )
+ \frac{1}{2 R_B \Rh F(R_B)} \left (k+\frac{3
\Rh^2}{\ell ^2} -
\frac{Q^2}{\Rh^2}\right )^2 \right ]
\ \ ,
\label{Tr}
\\
T_{Q^2}
&:=&
\frac{\partial T_H}{\partial \left ( Q^2\right )}
\nonumber
\\
&=&
\frac{1 }{4 \pi \sqrt{F(R_B)}}
\left[ -\frac{1}{\Rh^3} +
\frac{1}{2 R_B \Rh F(R_B)}
\left (
k+\frac{3 \Rh^2}{\ell ^2} - \frac{Q^2}{\Rh^2}\right )
\left ( \frac{1}{\Rh}
- \frac{1}{R_B} \right ) \right]
\ \ .
\label{Tq}
\end{eqnarray}
\end{mathletters}%
The range of the parameters is
\begin{mathletters}
\begin{eqnarray}
&&
\sqrt{\max ( 0, -\case{1}{3} k \ell^2 )} < \Rh < R_B
\ \ ,
\label{app-rh-ineq}
\\
&&
0 \leq Q^2 < \Rh^2
\left (\frac{3 \Rh^2}{\ell ^2} + k\right )
\ \ .
\label{app-q-ineq}
\end{eqnarray}
\end{mathletters}%
Here, (\ref{app-q-ineq}) and the leftmost inequality in (\ref{app-rh-ineq})
are the conditions for the existence of a nondegenerate
horizon. The rightmost inequality in (\ref{app-rh-ineq}) is the
condition that the black hole fit in the box.

We shall mainly discuss the sign of the heat capacities in the limit where
$R_B$ is taken to infinity while the parameters $\Rh$ and $Q^2$ remain in
some prescribed finite range. This means neglecting the second terms
in~(\ref{Trq}). In this limit, $T_{Q^2}$ is always negative, and
$T_{\Rh}$ is positive except when the following set of conditions holds:
\begin{eqnarray}
&&k=1
\ \ ,
\nonumber
\\
&&Q^2/\ell^2 < 1/36
\ \ ,
\nonumber
\\
&&{1\over6} \left( 1-\sqrt{1-36(Q/\ell)^2} \right)
< \Rh^2/\ell^2 <
{1\over6} \left( 1+\sqrt{1-36(Q/\ell)^2} \right)
\ \ .
\label{special-case}
\end{eqnarray}

Consider first~$C_Q$. With $X=Q$, we have $\left(dQ^2/d\Rh\right )_X=0$,
and the sign of $C_Q$ agrees with the sign of~$T_{\Rh}$. Hence, in the
limit $R_B\to\infty$, $C_Q$ is positive for all configurations except those
satisfying~(\ref{special-case}). As the second term in (\ref{Tr}) is
positive definite, taking $R_B$ finite would increase~$C_Q$, preserving the
stability for $k=0$ and $k=-1$ and widening the domain of stability for
$k=1$.

Consider next~$C_{\tilde{\Phi}_B}$. With $X={\tilde{\Phi}_B}$, we now
have\footnote{Note that in the limit $R_B\rightarrow \infty$ we have
$\left ( dQ^2/ d \Rh \right )_{\tilde{\Phi}_B}=\left (
dQ^2/ d\Rh\right )_{\phi _{\rm h}}$, where $\phi _{\rm h}:=Q\Rh^{-1}$ is
the quantity held fixed in the grand canonical ensemble in
section~\ref{sec:infinite-thermo}. This provides a check on the positivity
of the heat capacity $C_\phi$ discussed in
section~\ref{sec:infinite-thermo}.}
\begin{equation}
\left ( \frac{dQ^2}{d \Rh}\right )_{\tilde{\Phi}_B} =
\frac{2Q^2}{\Rh}\left [1 +  O (R_B^{-1}) \right ]
\ \ ,
\label{drhdqagain}
\end{equation}
and the denominator in (\ref{spec-heat2}) becomes
\begin{eqnarray}
T_{\Rh} + T_{Q^2} \left ( \frac{dQ^2}{d \Rh}\right )_{\tilde{\Phi}_B}
&=&
\frac{1}{4 \pi \sqrt{F(R_B)}}
\left\{ \frac{Q^2}{\Rh^4}
\left[ 1 +
\frac{\Rh^2}{Q^2}
\left (
\frac{3 \Rh^2}{\ell ^2}-k \right ) \right ] + O(R_B^{-1})\right \}
\ \ .
\label{denom}
\end{eqnarray}
In the limit $R_B\to\infty$, the expression in (\ref{denom}) is positive
definite for $k=0$ and $k=-1$. For $k=1$, however, it
becomes negative when the following set of conditions holds:
\begin{eqnarray}
&&k=1
\ \ ,
\nonumber
\\
&&
\Rh < \frac{\ell }{\sqrt{3}}
\ \ ,
\nonumber
\\
&&
Q^2 <
\Rh^2 \left( 1-\frac{3 \Rh^2}{\ell ^2} \right)
\ \ .
\end{eqnarray}
Thus, in the limit $R_B\to\infty$, $C_{\tilde{\Phi}_B}$ is positive definite
for $k=0$ and $k=-1$, but indefinite for $k=1$. For finite~$R_B$, it can be
verified that the terms omitted from (\ref{denom}) are positive definite:
$C_{\tilde{\Phi}_B}$ is positive for $k=0$ and $k=-1$ also with a finite
boundary, whereas for $k=1$, taking the boundary finite widens the domain
of stability.

Consider finally~$C_{\Phi _B}$. With~$X=\Phi _B$, we have
\begin{eqnarray}
\left ( \frac{dQ^2}{d\Rh}\right )_{\Phi _B}
&=&
Q^2\left ( k +
\frac{3 \Rh^2}{\ell ^2} -
\frac{Q^2}{\Rh^2}\right )
\left[
R_B\left ( k + \frac{R_B^2}{\ell ^2}\right )
- \Rh\left ( k + \frac{\Rh^2}{\ell ^2}\right )
\right]^{-1}
\nonumber
\\
&=&
\frac{Q^2 \ell^2}{R_B^3}
\left ( k + \frac{3 \Rh^2}{\ell ^2} -
\frac{Q^2}{\Rh^2}\right )
\left[ 1+ O(R_B^{-1})\right ]
\ \ ,
\end{eqnarray}
and the denominator in (\ref{spec-heat2}) becomes
\begin{equation}
T_{\Rh} + T_{Q^2}\left ( \frac{dQ^2}{d
\Rh}\right )_{\Phi _B} =
\frac{1}{ \Rh^2}\left (-k + \frac{3 \Rh^2}{\ell ^2}
+ \frac{3 Q^2}{\Rh^2}\right)
\left[ 1+ O(R_B^{-1})\right ]
\ \ ,
\label{denom2}
\end{equation}
In the limit $R_B\to\infty$, the expression in (\ref{denom2}) is positive
definite for $k=0$ and $k=-1$. For $k=1$,
it is negative when the conditions (\ref{special-case}) hold. Hence, in the
limit $R_B\to\infty$, $C_{\Phi _B}$ is positive definite for
$k=0$ and $k=-1$ and indefinite for $k=1$.

These results show that for $k=0$ and $k=-1$, the heat
capacities $C_Q$, $C_{\tilde{\Phi}_B}$, and~$C_{\Phi _B}$ are positive
definite in the limit $R_B\to\infty$. There exist, however,
choices for $X$ such that $C_X$ can be negative for $k=0$ and $k=-1$:
from~(\ref{Trq}), it is seen that this happens whenever
$\left (d Q^2/d\Rh\right )_X > \Rh\left (-k +3 \Rh^2/\ell^2 + 3
Q^2/\Rh^2\right )$. Saturating this inequality
corresponds to $X=F'(\Rh)$, which is equivalent to holding the renormalized
temperature $T_\infty$ constant.

The signs of our three heat capacities, in the limit of a large box, may be
summarized in the following table:
\begin{equation}
\begin{array}{c|ccc}
& C_Q & C_{\tilde{\Phi}_B} & C_{\Phi _B}\\ \hline
k=-1 & + & + & + \\
k=0 & + & + & + \\
k=1 & \pm &  \pm & \pm \end{array}
\label{spec-heat-list2}
\end{equation}

\end{document}